\documentclass[fleqn, 11pt]{wlscirep}
\title{Deep Learning-Assisted Weak Beam Identification in Dark-Field X-ray Microscopy} 
\usepackage{gensymb}
\usepackage{multirow}
\usepackage{xcolor}
\def\tablename{Table}%
\usepackage{float}
\usepackage{amsmath} 
\usepackage{epstopdf}           %
\usepackage{flushend}           %
\usepackage{ragged2e}%

\usepackage{lmodern}
\usepackage[textwidth=1.8cm]{todonotes}  

\usepackage{lineno}
\usepackage{xr}
\usepackage[figuresright]{rotating}
\usepackage{soul}

\usepackage{hyperref}
\hypersetup{
    colorlinks=true,
    linkcolor=blue,
    filecolor=magenta,      
    urlcolor=cyan,
    pdftitle={Overleaf Example},
    pdfpagemode=FullScreen,
    }

\usepackage{subfigure}

\author[1,2 ]{A. Benhadjira}
\author[1]{C. Detlefs}
\author[3]{S. Borgi}
\author[1,2]{V. Favre-Nicolin}
\author[1]{C. Yildirim*} 
\affil[1]{European Synchrotron Radiation Facility, 71 Avenue des Martyrs, CS40220, 38043 Grenoble Cedex 9, France.}
\affil[2]{Univ. Grenoble - Alpes, 38000 Grenoble, France}
\affil[3]{Technical University of Denmark, Department of Physics, Lyngby , Denmark}

\affil[*]{Corresponding author: can.yildirim@esrf.fr}


\begin{abstract}

Dislocations control the mechanical behavior of crystalline materials, yet their quantitative characterization in bulk has remained elusive. Transmission Electron Microscopy provides atomic-scale resolution but is restricted to thin foils, limiting relevance to structural performance. Dark-field X-ray microscopy (DFXM) has recently opened access to three-dimensional, non-destructive imaging of dislocations in macroscopic crystals. A critical bottleneck, however, is the reliable identification of weak- versus strong-beam conditions. Weak-beam imaging enhances dislocation contrast, while strong-beam conditions are dominated by multiple scattering and obscure interpretation. Current practice depends on manual classification by specialists, which is subjective, slow, and incompatible with the scale of modern experiments. Here, we introduce a deep learning framework that automates this task using a lightweight convolutional neural network trained on small, hand-labeled datasets. By enabling robust, rapid, and scalable identification of imaging conditions, this approach supports scalable DFXM analysis, unlocking statistically significant studies of dislocation dynamics in bulk materials.

\end{abstract}
\begin{document}
\flushbottom
\maketitle

\section*{Introduction}
Dislocations are fundamental to understanding both the mechanical and functional behavior of crystalline materials. As the primary carriers of plastic deformation, they mediate irreversible atomic motion within the lattice, dictating how materials yield, harden, and ultimately fail under applied stress \cite{Taylor1934, polanyi1934grid, anderson2017theory}. Their role extends well beyond mechanical response: dislocations also shape key functional properties, including electrical conductivity, magnetic behavior, and atomic diffusion \cite{Reiche2016,brown1941effect,Hughes2021,Broudy1963}. Central to most experimental methods for probing dislocations is their angular sensitivity. In an ideal crystal, lattice planes are perfectly periodic, giving rise to sharp diffraction of incident X-rays or electrons at well-defined angles. Dislocations perturb this periodicity through their long-range strain fields, locally tilting lattice planes by fractions of a degree and shifting the diffraction condition by only milliradians. Owing to the high selectivity of diffraction, such subtle distortions generate measurable contrast, enabling direct imaging of dislocations.

Visualizing and characterizing dislocations across relevant length scales is essential for uncovering the mechanisms that govern deformation. Transmission electron microscopy (TEM) has played a foundational role in this effort, providing near-atomic resolution images that reveal dislocation core structures and interactions \cite{Williams2009,Meng2021}. However, the requirement for electron-transparent samples, typically limited to thicknesses below a few hundred nanometers, restricts TEM’s ability to probe three-dimensional dislocation networks within bulk crystals, especially those with low dislocation densities such as annealed single crystals. Additionally, for systems like semiconductors and ceramics, where dislocation densities are inherently low, the limited field of view in TEM poses a challenge for capturing statistically significant defect populations.

To overcome these limitations, dark-field X-ray microscopy (DFXM) has emerged as a powerful, non-destructive method for imaging deeply embedded dislocations under realistic conditions \cite{Simons2015, Poulsen2021, Jakobsen2019, Yildirim2023}. Using high-energy synchrotron X-rays and objective-based imaging in Bragg diffraction geometry, DFXM enables real-space visualization of lattice distortions with sub-micrometer spatial resolution and angular sensitivity on the order of $10^{-4}$ radians. Conceptually similar to dark-field TEM but with greater penetration depth and tunable magnification, DFXM achieves resolution limits around 50 nm. Notably, Jakobsen et al.\ \cite{Jakobsen2019} demonstrated the use of weak-beam (WB) contrast in DFXM, adapting a concept long used in TEM \cite{Cockayne1969}, to selectively image regions with high lattice distortion by recording at the tails of the rocking curve. This approach highlights dislocation lines with high specificity and clarity, providing detailed 2D projections of dislocation structures.

\begin{figure}[t]
    \centering
    \includegraphics[width=\textwidth]{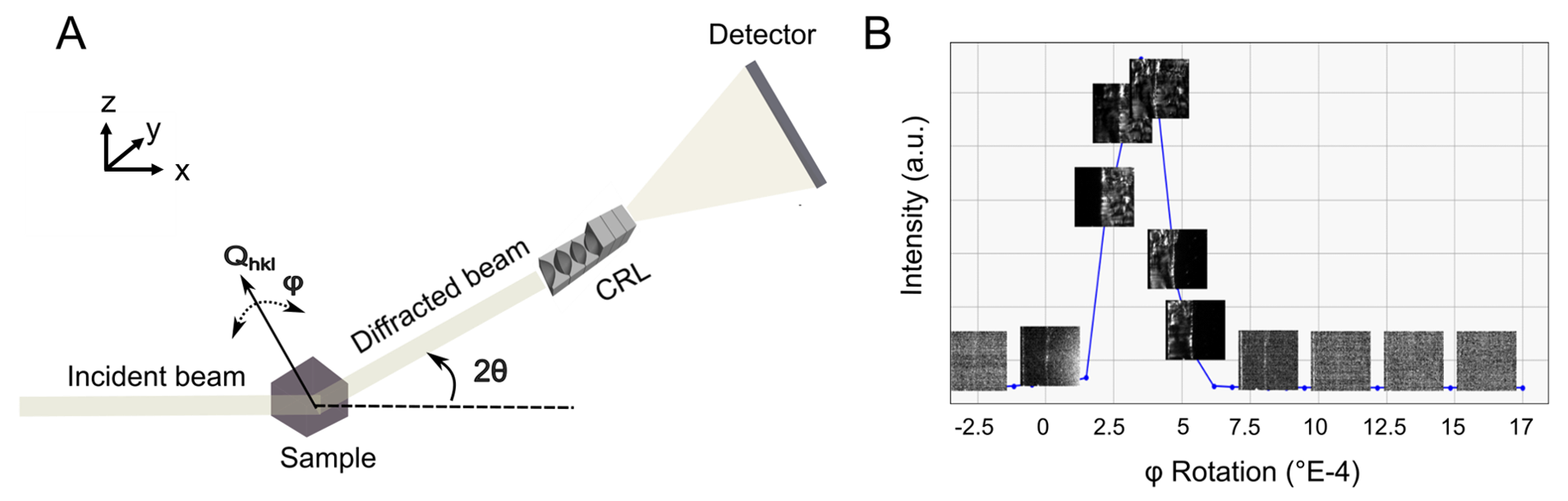}
    \vspace{-5mm}
    \caption{\justifying \textbf{Schematic of a Dark-Field X-ray Microscopy (DFXM) experiment.} 
    (A) A line-focused X-ray beam (approximately 500 nm in height in $z$ direction and a couple of hundreds of microns wide in $y$ direction, illuminates a layer of the crystal. The diffracted beam at the scattering angle \(2\theta\) is selected by a compound refractive lens (CRL) system and imaged onto a 2D detector \cite{Isern2024}. Rotating the sample around the \(\phi\)-axis left handed rotation around y axis enables collection of a rocking curve, capturing lattice distortions and rotations about the scattering vector \(\mathbf{Q}\). 
    (B) A representative rocking curve obtained from fine angular steps. Each point corresponds to an image recorded at a discrete ($\phi$) angle. The curve's tails reflect WB conditions, sensitive to dislocations and strain, while the peak corresponds to SB conditions, often showing dynamical diffraction fringes.}
    \label{dfxm}
\end{figure}

By mapping the local lattice orientation variations and elastic strain around a given diffraction vector, DFXM acquires spatially-resolved high-dimensional (up to 6D in static measurements) datasets that capture long-range strain fields and the spatial arrangement of dislocations in three dimensions. 
Datasets from subsequent virtual sections can be stacked to reconstruct volumetric images of defects, enabling \emph{in situ} studies of dislocation network evolution under mechanical loading \cite{Simons2015, Dresselhaus-Marais2021, Frankus2025}. 
However, DFXM data analysis remains data intensive and requires significant supervision. In particular, this has become more prominent with the advent of extremely brilliant synchrotron sources (EBS) \cite{Raimondi2023, Bruno2024} which allow mapping an entire grain within minutes \cite{Yildirim2025}. 

Here we address a specific problem relevant for samples of high crystal quality, e.g.~samples where individual dislocations within dislocation networks such as coherent small angle grain boundaries can be resolved.
In such samples, strong multiple scattering occurs, and the dynamical theory of x-ray diffraction has to be applied in order to simulate the topographs observed near the maximum of the rocking curve \cite{Authier2003,Epelboin1979}.
In the tails of the rocking curve, however, multiple scattering is negligible and the simpler kinematic theory of x-ray diffraction can be applied. This is called the WB condition.
Therefore, the typical approach to analyzing rocking curve imaging (RCI) or DFXM data is to first segment the scanning data into weak- and strong-beam (SB) images, and then analyze and interpret the weak-beam images.
The aim of this work is to provide a Machine Learning (ML) workflow for the automatic identification of weak-beam images in DFXM rocking curves. 
In subsequent data analysis steps, these weak beam images can the be used for the automatic identification of dislocations and other localized crystal defects.

Conventional rocking curves analysis methods rely on manual interpretation or unsupervised approaches such as principal component analysis \cite{GarrigaFerrer2023} or Gram-Schmidt orthogonalization \cite{Huang2023} to identify the weak-beam condition and visualize strain fields near defect cores. However, these methods are frequently time-consuming and lack generalizability across different datasets and material systems. As a result, there is a pressing need for more automated, robust, and flexible analysis pipelines.

Machine Learning (ML) algorithms can offer a promising solution for the above-mentioned challenges. In recent years, ML methods have been increasingly applied to diverse materials science problems \cite{Butler2018,Schmidt2019,Ramprasad2017,Xie2018,Jha2018}. The success of deep learning in computer vision has inspired materials scientists to adopt these techniques for tasks such as automatic classification and segmentation of microstructural features. ML approaches have shown promise in uncovering relationships between microscopic structures and material properties, supporting both theoretical studies and simulations. For example, Zhang et al.\ \cite{Zhang2019} used convolutional neural networks (CNNs) to extract dislocation microstructures from electron backscatter diffraction (EBSD) data, while Yassar et al.\ \cite{Yassar2010} employed artificial neural networks (ANNs) to predict the flow stress of micropillars using dislocation density and model size as input features. Similarly, Tao et al.\ \cite{Tao2024} applied various ML models to predict micropillar compression behavior based on discrete dislocation dynamics (DDD) simulations, and Hiemer et al.\ \cite{Hiemer2023} used Kernel Ridge Regression (KRR) to link plasticity with dislocation characteristics, analyzing the effects of strain rate, dislocation density, and strengthening mechanisms in FCC metals.

Despite their potential, deep neural networks often require large numbers of parameters and significant computational resources. To address this, lightweight network architectures \cite{mobilenet, lightweightsur} have been developed to maintain accuracy while reducing model complexity and computational cost. Such models are well-suited for applications with limited computational resources. In the context of DFXM, the WB condition is not typically normally distributed and can appear irregularly within individual rocking curve frames. Moreover, not all data exhibit clearly separable weak- and strong-beam features, and user interpretation often plays a role in manual analysis. To address these challenges, we developed a patch-based ML workflow that reduces DFXM dataset size while automating the classification of diffraction contrast. This approach enables consistent identification of WB and SB conditions, streamlining defect analysis and reducing subjective bias. Our method offers a faster and more reliable route for analyzing DFXM dislocation data, and can be extended to related techniques that rely on similar contrast mechanisms, such as X-ray topography \cite{Caliste2021} and transmission electron microscopy \cite{WilliamsCarter1996}.
\begin{figure}
    \centering
    \includegraphics[width=\linewidth]{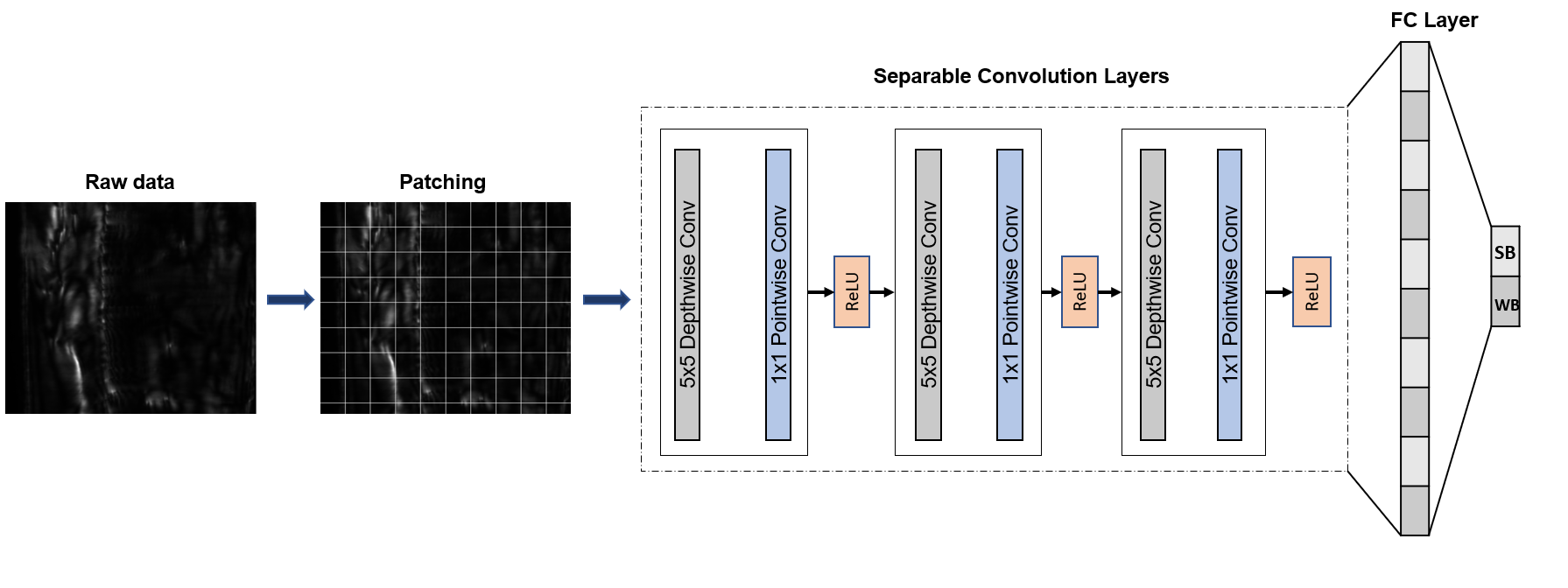}
    \caption{\justifying \textbf{Deep learning architecture for DFXM image patch classification.} 
    Raw DFXM images are divided into smaller patches that are independently analyzed using a lightweight convolutional neural network. Each patch is passed through a sequence of depthwise and pointwise separable convolution layers with ReLU activation functions, enabling efficient feature extraction. The resulting features are then fed into a fully connected (FC) layer that classifies each patch as corresponding to either a weak-beam or strong-beam diffraction condition.}
    \label{cnn}
\end{figure}

\section*{Results}

\subsection*{Deep learning model}

Herein, we aim to develop automated methods for reducing and analyzing data from three-dimensional dislocation imaging using DFXM. 
A typical setup is illustrated in Figure~\ref{dfxm}A, where a single crystal or grain of interest is aligned through the objective with a given $(hkl)$ reflection. 
An incident X-ray beam illuminates the sample, and the diffracted beam is selected at the Bragg angle $2\theta$ using a compound refractive lens (CRL) system and directed to a 2D detector. 
For typical 3D scans, a line-focused incident beam is used, and the sample is translated along the laboratory $z$-axis to sequentially probe 2D virtual slices through the volume. 
At each $z$-layer, the sample is rotated around an axis perpendicular to the incident beam, typically the $\phi$-axis \cite{Poulsen2017}, to collect a rocking curve, with DFXM images acquired at each angular step, as shown in Figure~\ref{dfxm}B. 
Traditionally, WB images, those located at the tails of the rocking curve, are manually identified to enhance dislocation contrast. 
Accurate and reproducible reconstruction of the three-dimensional dislocation structure requires robust identification of appropriate weak-beam conditions, motivating the need for automated analysis tools.
\begin{figure}[t]
    \centering
    \includegraphics[width=\linewidth]{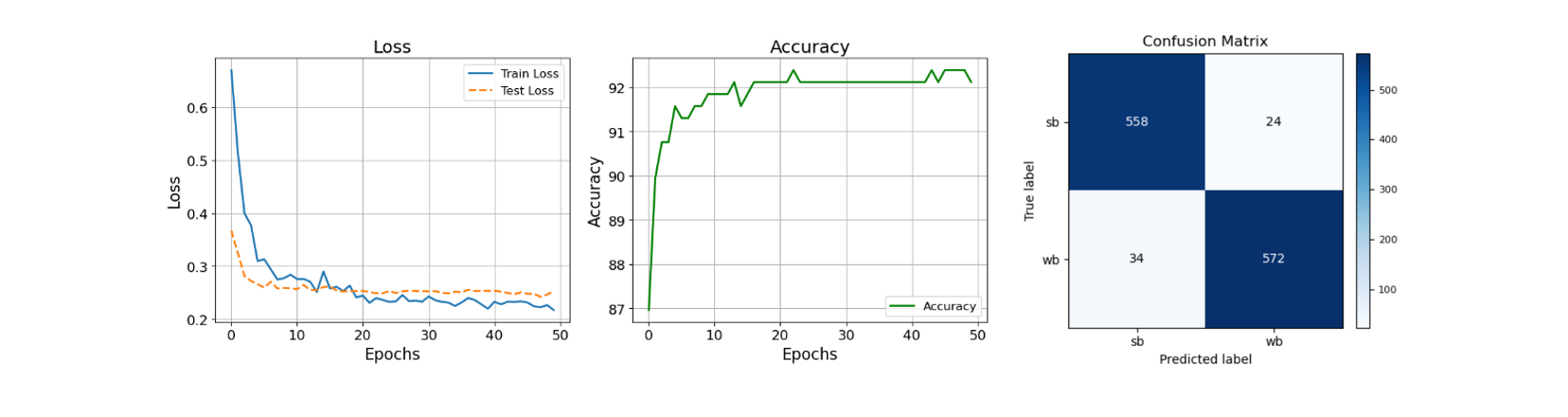}
    \caption{\justifying \textbf{Performance metrics of the classification model.} 
    From left to right: (1) Training and test loss plotted over 50 epochs, showing a decreasing trend and convergence, indicating effective learning and minimal overfitting. (2) Accuracy curves for both training and test sets, demonstrating stable and high performance throughout training. (3) Confusion matrix summarizing the validation dataset classification results, highlighting the model’s ability to distinguish between weak-beam and strong-beam conditions.}
    \label{loss}
\end{figure}

To preserve all relevant information and enable reliable 3D reconstruction, we propose a patch-based approach for the localized identification of weak-beam conditions in each rocking layer. This method allows for small-patch-wise classification of beam conditions. Ground truth labels for training were manually generated by sampling representative frames from the dataset and annotating them according to their corresponding beam condition.

The complete 3D dataset analyzed in this work (originally reported in Ref.~\citeonline{Yildirim2023}) comprises 9311 frames, each containing approximately 4 million pixels. From this volume, six representative frames, 3 weak beam  and 3 strong beam (0.064$\%$ of the total dataset) were selected for manual annotation. Labeling was performed on a per-patch basis rather than per full frame, and each patch was classified as either WB or SB, resulting in a binary classification task. Because local lattice curvature and strain may cause spatial variations in diffraction conditions within a single projection, particular care was taken during annotation. Only patches exhibiting clearly distinguishable WB or SB contrast were selected for training, while ambiguous or transitional regions were excluded to avoid label noise. The selected frames span representative diffraction conditions across the rocking curve, ensuring that the training data captures realistic contrast variability while maintaining physically well-defined labels.

For our classification task of interest, we utilize a lightweight convolutional neural network (LCNN) based on a depthwise separable architecture \cite{pointwise, depthwise}. As depicted in Figure 2, the model operates on relatively small image patches $64 \times 64$ pixels extracted from raw DFXM data.

The subdivision of full-sized raw images into smaller patches offers several advantages that enhance both the training efficiency and robustness of the model. By operating on smaller datasets, the LCNN reduces the number of variables, thereby improving computational efficiency. Additionally, in samples with structural complexity, such as strong curvature or subgrain boundaries, different regions of the image may exhibit the weak-beam condition at varying orientations. Using patches that are comparable in size to the characteristic features helps to avoid inconsistencies across the image. This localized approach also enables more effective use of labeled data; a single hand-labeled full-sized image can yield many smaller training patches (up to 1024 patches), significantly increasing the volume of training data without additional labeling effort. Once the model processes each patch, the segmented outputs can be reassembled into full-sized images for interpretation.

Each patch then undergoes a sequence of depthwise separable convolutional layers, beginning with 5×5 depthwise convolutions that extract spatial features within individual channels, followed by 1×1 pointwise convolutions that integrate information across channels. These features are projected into a 32-dimensional latent space, with non-linearities introduced via rectified linear unit (ReLU) activations to enhance representational capacity. The sequence is repeated multiple times to capture complex image characteristics, and the final representation is passed to a fully connected layer that classifies each patch as exhibiting either strong-beam or weak-beam conditions. To improve generalization and prevent overfitting, dropout regularization \cite{JMLR} with a probability of 0.15 is applied during training.

\subsection*{Model evaluation}
The data used in this study originate from two previously published and publicly available DFXM datasets corresponding to two separate samples: Sample 1 reported in Ref.~\citeonline{Yildirim2023} and Sample 2 reported in Ref.~\citeonline{Borgi2025}. Detailed descriptions of the samples and experimental conditions are provided in the Data Acquisition section. The model was trained using the dataset from Sample 1. Each rocking step frame had a resolution of 2160×2560 pixels, which was divided into non-overlapping 64×64 patches. Labeling was performed manually at the image level, meaning that entire images were classified as either SB or WB conditions. Based on this, 3 WB images and 3 SB images were randomly selected, resulting in 6144 patches for each beam condition. The ground truth dataset was split into 80\% for training and 20\% for testing and $\sim$1300 validation patches. All patches were log-scaled to enhance feature contrast and were passed through the neural network in mini-batches during training to ensure efficient learning and generalization.

The model is trained for 50 epochs, converging around epoch 30, as shown in Figure \ref{loss}. Both training and testing losses decrease steadily, with no significant divergence, indicating effective model generalization. The use of dropout further enhances model robustness by deactivating a subset of neurons during training epochs. Classification performance is evaluated using accuracy, defined as the proportion of correctly classified samples relative to the total dataset size:

\begin{equation} \text{Accuracy} = \frac{\text{TP} + \text{TN}}{\text{TP+FP+TN+FN}} \end{equation}

where TP and TN denote true positives and true negatives, respectively, and FP and FN represent false positives and false negatives.Additionally, the final epoch confusion matrix provides further insight, showing that the majority of the validation set was correctly classified. However, we observe that 10 additional WB patches were misclassified. A likely explanation is related to intensity variations under the SB condition. SB images often contain dynamical fringes and local intensity fluctuations. In regions where the WB contrast is weak or transitional (i.e., leaning toward SB), these variations can cause a WB patch to resemble SB at the patch level. This can bias the prediction toward SB and lead to the observed minor asymmetry. That said, the effect is subtle and mainly occurs in borderline regions rather than in clearly defined dislocation segments.

To benchmark our LCNN and evaluate the gains in accuracy relative to computational cost, we compared it with ResNet18 \cite{res} and VGG16 \cite{vgg}, which are lightweight variants of the ResNet and VGG families. Table \ref{tab:model_comparison} summarizes their respective performance in terms of parameter count, training time, accuracy, and deployment time for labeling 300 rocking layers. \textcolor{blue}The models were implemented using PyTorch, a Python deep learning library, and trained for 50 epochs with batches of 32 patches at a time on NVIDIA Tesla V100-SXM2-32GB GPUs.

\begin{table}[h]
    \centering
    \resizebox{\textwidth}{!}{%
    \begin{tabular}{lr@{}lr@{}lr@{}lr@{}l}
        \toprule
        \textbf{Model} 
        & \multicolumn{2}{c}{\textbf{Trainable Parameters}} 
        & \multicolumn{2}{c}{\textbf{Training Time (S)}} 
        & \multicolumn{2}{c}{\textbf{Accuracy (\%)}} 
        & \multicolumn{2}{c}{\textbf{Deployment Time (h)}} \\
        \midrule
        LCNN     & 879&K   & 23&.87   & 92&.70  & 0&.41  \\
        ResNet18 & 11&M    & 75&.07   & 97&.13  & 1&.32 \\
        VGG16    & 134&M   & 208&.09  & 89&.10  & 1&.55  \\
        \bottomrule
    \end{tabular}%
    }
    \caption{Performance comparison of LCNN, ResNet18, and VGG16 in terms of model size, training time, accuracy, and deployment time.}
    \label{tab:model_comparison}
\end{table}

\begin{figure}[h]
    \centering
    \includegraphics[width=0.6\linewidth]{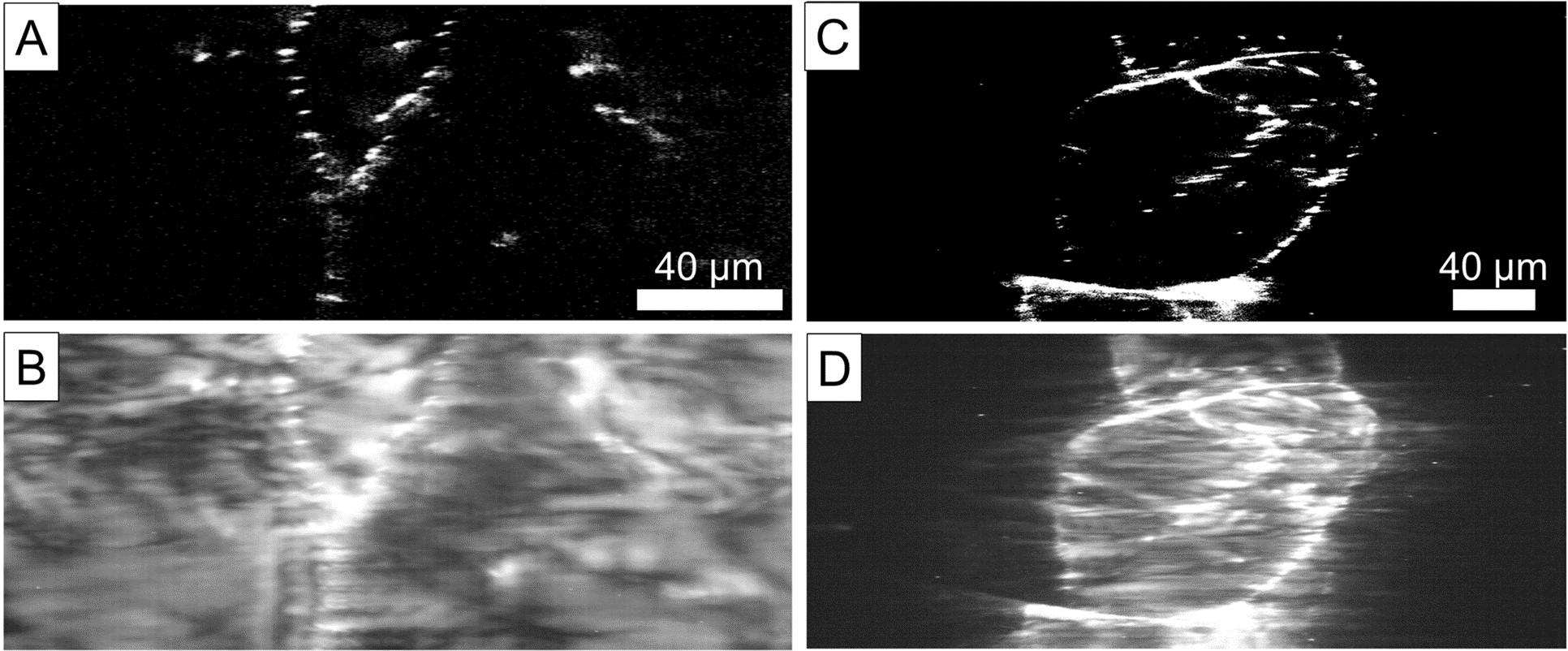}
    \caption{\justifying \textbf{Classification of weak- and strong-beam conditions using LCNN.} 
(A) and (B) show diffraction contrast maps from the training dataset, automatically classified by our lightweight model as WB and SB conditions, respectively. In both cases, dislocation boundaries and branching structures are visible, and individual dislocations can be resolved. 
(C) and (D) show predictions on an independent experimental dataset that is different from the training dataset (annealed Al, probed at the \((111)\) reflection, 17~keV), where the model correctly identifies weak-beam and strong-beam conditions, respectively \cite{Borgi2025}. Training data are from Ref.~\citeonline{Yildirim2023}. The scale bar in (A) applies also to (B), and the scale bar in (C) applies also to (D).}

    \label{validation}
   
\end{figure}

While ResNet18 achieves the highest classification accuracy, our LCNN provides a solid balance between performance and efficiency. Due to its lightweight architecture, it has significantly fewer parameters and trains faster, making it well-suited for real-time applications and environments with limited computational resources. 
Additionally, the patch size impacts both training/inference time and accuracy, therefore it is important for users to choose a patch size appropriate to their dataset. Based on our experiments, a patch size of $64 \times 64$ pixels offers the best trade-off between accuracy and computational cost.

Although the field of view and pixel size differ between the 10$\times$ and 2$\times$ objectives, the weak-beam contrast associated with individual dislocations remains comparable in spatial extent, as illustrated in Fig.~\ref{validation}(a,b) for 10$\times$ and Fig.~\ref{validation}(c,d) for 2$\times$. Therefore, the selected 64×64 patch size does not require systematic rescaling with magnification. Only in cases of significantly higher dislocation density, where multiple strain fields overlap within a single patch, would a smaller patch size be advantageous.

\subsection*{Discussion}
After training, we proceed by assigning our LCNN to label the entire rocking curves by identifying WB regions, which are particularly of interest for dislocation mapping. To enhance defect visibility, frames corresponding to WB-labeled patches are extracted and combined into an integrated image. This approach excludes SB contributions dominated by dynamical diffraction. Figure~\ref{validation} shows the integrated patches under both strong and weak beam conditions. Here, integrated WB and SB images are constructed by summing all rocking-curve patches identified as WB or SB, respectively, thereby producing a single representative WB image and SB image for further quantitative analysis and 3D reconstruction. Postprocessing is then applied to improve image contrast in the integrated WB images:

\begin{itemize}
\item Background subtraction: The mean intensity plus one standard deviation is subtracted from each pixel.
\item Normalization: Pixel values are scaled by the maximum intensity in the image.
\item A fixed intensity threshold is applied to generate a binary map, highlighting potential defect regions.
\end{itemize}

Figure~\ref{validation} (A) and (B) shows LCNN-labeled diffraction contrast maps under WB and SB conditions after post-processing. WB conditions enhance dislocation contrast, making dislocation boundaries and branching structures more distinct. Furthermore, Fig. ~\ref{validation} (C) and (D) demonstrate model predictions on previously unseen experimental data, where the LCNN correctly distinguishes WB from SB. Finally, the WB-labeled rocking layers are then assembled to reconstruct three-dimensional dislocation networks for spatially resolved analysis of defect structures.

\begin{figure}[t]
    \centering
    \includegraphics[width=\linewidth]{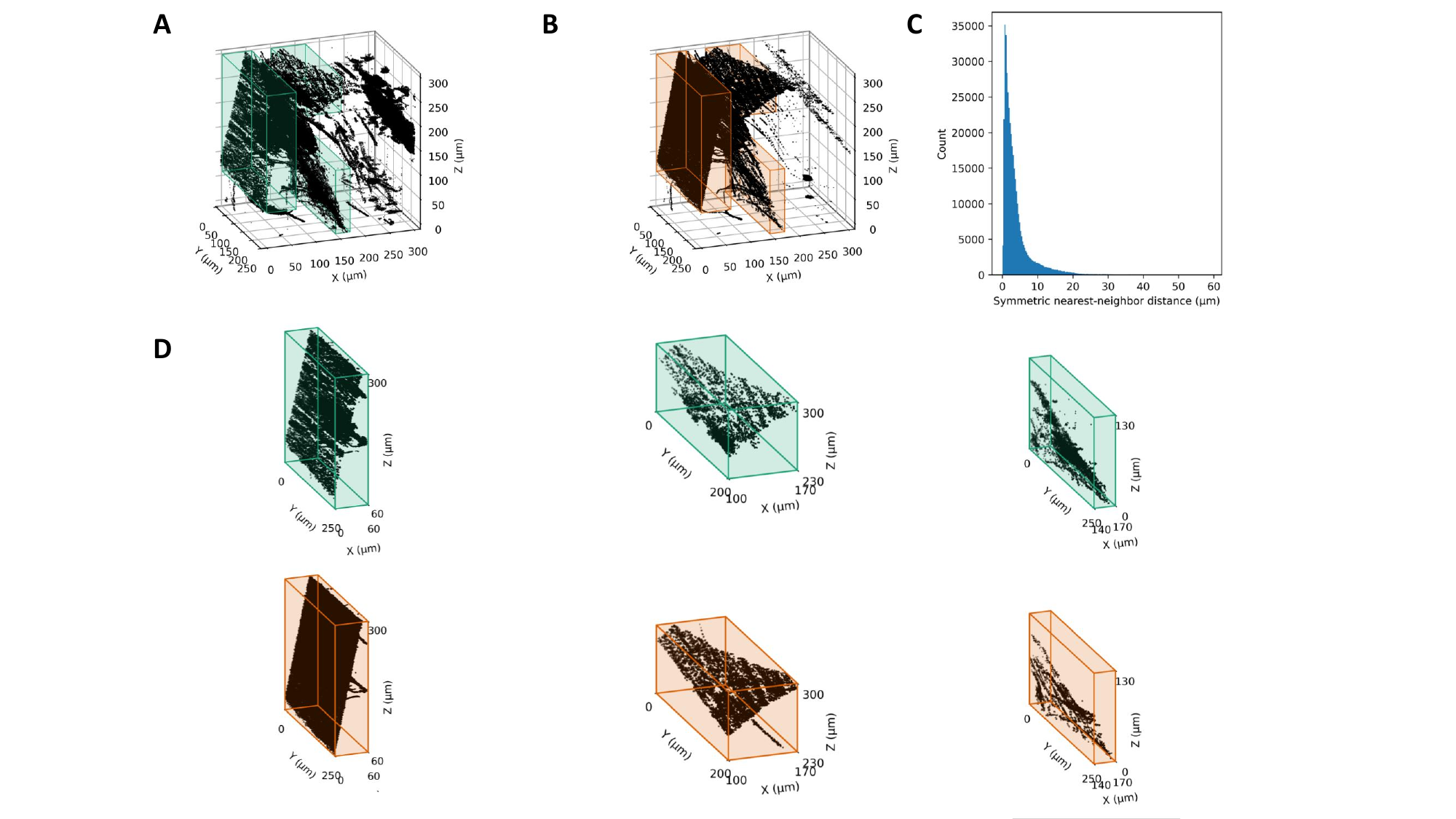}
   \caption{\justifying \textbf{Comparison of dislocation structure reconstruction using automated and manual methods.} 
(A) 3D reconstruction of individual dislocations and dislocation boundaries using our fully automated machine learning workflow based on weak-beam identification.  
(B) Corresponding 3D reconstruction using manual intensity thresholding from the published experimental dataset~\cite{Yildirim2023}.  
(C) Distribution of symmetric nearest-neighbor distances between the automated and manually reconstructed dislocation networks, quantifying the spatial agreement between the two approaches.  
(D) Representative dislocation boundaries ROIs from the reconstructed volumes (automated: green; manual: orange), demonstrating close structural correspondence across multiple 
Despite relying on significantly less data and requiring no human intervention. }
\label{3d}
\end{figure}

To place this work in context, it is instructive to compare our LCNN-based workflow with other relevant approaches for weak-beam condition identification in DFXM. 
For instance, Huang et al \cite{Huang2023} proposed a method similar to principal component analysis (PCA) that uses Gram-Schmidt orthogonalization (GSO) instead to decompose rocking curve data into three orthogonal components corresponding to the SB condition and the WB conditions on either side of the rocking curve. Unlike the data-driven approach employed here, their method does not rely on any training data but instead defines physically meaningful basis functions from manually selected regions of interest. The main advantage of such a physics-based orthogonalization is its interpretability and its grounding in the underlying diffraction contrast mechanisms. However, this also makes it more dependent on careful manual input and less straightforward to implement in an automated or real-time setting, particularly for large, heterogeneous datasets. 

In the GSO formulation, the image stack is represented as a linear combination of three orthogonal basis states corresponding to WB-left, WB-right, and SB conditions. However, this orthogonality assumption is not generally satisfied in experimental datasets. In practice, WB-left and WB-right conditions often exhibit asymmetric contrast variations that depend on the local microstructure and diffraction condition, which can lead to incomplete separation and residual SB contamination in the extracted WB components. Furthermore, both PCA and GSO require manual selection of regions of interest to optimize the decomposition, typically repeated for individual 2D layers within a 3D scan. The outcome therefore becomes sample and operator dependent, and the procedure is time-intensive. In contrast, the LCNN approach is trained using image patches extracted from randomly selected frames within a dataset and can then be applied to the entire volume without further manual intervention. This ensures consistent treatment across the dataset and significantly reduces user bias. When reproducing the GSO analysis reported by Huang et al. \cite{Huang2023}, we observe that the GSO-derived WB component remains partially contaminated by residual SB fringes, whereas the LCNN output isolates dislocation contrast with substantially reduced SB background contributions.  The LLCN presented in this work therefore enables automated, patch-level classification of beam conditions with minimal computational cost, supporting high-throughput processing of large DFXM volumes. The trade-off is that the LCNN requires representative training data and may need retraining when applied to material systems with substantially different contrast characteristics.

Together, these two strategies illustrate complementary approaches for achieving more efficient and scalable analysis of DFXM datasets. The LCNN provides a rapid and robust solution for weak-beam identification under diverse conditions, and its model-agnostic design makes it a promising candidate for application in other imaging modalities such as TEM and X-ray topography. Moreover, we emphasize that while the model may occasionally mislabel rocking curves from unfamiliar samples, the same workflow remains valid: by manually labeling a small subset of the new dataset and retraining the model, it can be effectively adapted to the specific features of interest.

As a final validation of our approach, we reproduced the three-dimensional dislocation structures previously published in Yildirim et al \cite{Yildirim2023}. Figure~\ref{3d} presents a direct comparison between the reconstruction reported by Yildirim et al., which is based on intensity thresholding applied to each individual 2D layer, and the reconstruction obtained using the LCNN approach proposed in this work. Fig.~\ref{3d}(A) shows the 3D reconstruction obtained through our fully automated workflow, whereas Fig.~\ref{3d}(B) corresponds to the manually segmented result from the original study after intensity thresholding. In Fig.~\ref{3d}(A), several individual dislocations are visible, which can be attributed to the integrated weak-beam approach that captures dislocation contrast under different weak-beam conditions throughout the rocking curve. At first glance, both approaches show similar features in the 3D volume in terms of dislocation boundaries. The visual similarity between the two volumes is further quantified in Fig.~\ref{3d}(C), which reports the distribution of symmetric nearest-neighbor distances between the reconstructed dislocation networks, with a mean error of approximately 2~$\mu$m, indicating close spatial agreement. Fig.~\ref{3d}(D) displays representative dislocation boundaries extracted from both datasets, highlighting the consistent structural features recovered across multiple ROIs. Although the dislocation boundaries are reliably reconstructed, minor artifacts are also observed in the automated result, likely introduced by the patching method. Despite using approximately 90$\%$ less data and requiring no human intervention,The automated method shows good agreement of $77\%$, corresponding to 5.4 million dislocation voxels compared to 4.2 million in the published results.This capability is particularly relevant for time-resolved experiments at advanced synchrotron and free-electron laser sources, such as ESRF-EBS and XFELs, where large volumes of data are acquired rapidly.

To better understand the distinction between WB and SB conditions, we evaluated our model across multiple datasets. While dislocation features are generally most evident under WB imaging conditions, lattice rotations introduced by dislocation boundaries frequently cause DFXM images to contain mixed WB and SB contributions within different regions of the same sample. As a result, in several rocking-curve positions, the intensity distributions of WB and SB frames partially overlap, particularly in transitional angular regions. This overlap complicates the interpretation of dislocation contrast and makes the separation between beam conditions ambiguous. Such ambiguity renders simple intensity threshold–based methods unreliable. Across the full dynamic range of detector intensities, variations in sensitivity to WB signal can introduce parasitic effects that further obscure the distinction between conditions. Consequently, manually selected WB frames may not be representative of the full scan range and can bias the reconstructed dislocation population.

This challenge is reflected quantitatively in the 3D reconstruction shown in Fig.~\ref{3d}. Within the same region of interest, conventional intensity thresholding yields approximately 4.2 million dislocation voxels (as reported in Ref.~\cite{Yildirim2023}), whereas the LCNN-based reconstruction identifies approximately 5.4 million voxels corresponding to an increase of about 25\% in reconstructed dislocation volume. The improved completeness arises from integrating information across the full rocking curve rather than relying on a single global intensity criterion. These results show the advantage of our data-driven approach, which captures subtle, context-dependent features beyond raw intensity values and provides a scalable solution for dislocation imaging across diverse materials systems and experimental configurations.

\section*{Conclusion and Outlook}

In this work, we presented an automated and computationally efficient framework for analyzing DFXM rocking curves using lightweight convolutional neural networks. Our model enables accurate and scalable identification of weak-beam conditions crucial for interpreting diffraction contrast and visualizing dislocations in crystalline materials. This approach substantially reduces manual intervention and enables efficient three-dimensional dislocation mapping in bulk crystalline materials.

Looking ahead, we aim to extend this method for real-time, on-the-fly data reduction during DFXM experiments, thereby streamlining acquisition and enabling adaptive imaging strategies. Concurrently, we are developing models to infer dislocation types and Burgers vectors by training neural networks on synthetic datasets generated from geometrical optics simulations \cite{Poulsen2021,Borgi2024}. These advancements will contribute toward a more quantitative, interpretable, and fully automated framework for defect characterization in crystalline materials.
The methodology presented here is also broadly applicable to other imaging modalities where WB contrast plays a critical role, including X-ray topography, white-beam Laue microscopy, and transmission electron microscopy.

\section*{Methodology}
\subsection*{Weak beam dark field microscopy}

Weak Beam Dark Field (WBDF) microscopy \cite{Williams2009} is an imaging technique used to enhance contrast from crystalline defects such as dislocations and strain fields \cite{Dovidenko1999,Lin2024}. By operating slightly off the exact Bragg condition \cite{Cockayne1972}, WBDF suppresses background contrast and amplifies strain-induced intensity variations. This makes it particularly effective for resolving nanoscale structural distortions and has been widely applied in electron microscopy to study dislocation contrast and dynamical effects \cite{Journot2023,peng2022dislocation}. When implemented with high-brilliance synchrotron sources, WBDF X-ray microscopy further enables non-destructive characterization of bulk materials, providing insight into strain distributions, phase transformations, and defect structures in metals, semiconductors, and related systems.

\subsection*{Data acquisition}
The data analyzed in this work originates from DFXM experiments conducted on two independent single-crystal aluminum samples during separate beam times at Beamline ID06-HXM (now ID03) of the European Synchrotron Radiation Facility (ESRF) \cite{Kutsal2019,Isern2024}.  The first dataset corresponds to the study reported in Yildirim et al.~\cite{Yildirim2023}, from which the full 3D dislocation reconstruction presented in Fig.~\ref{validation}(a-b) and Fig.\ref{3d} is derived. We refer to this as Sample 1. In this case, the (002) Bragg reflection was mapped. The second dataset corresponds to the measurements reported in Borgi et al.~\cite{Borgi2025}, Fig. \ref{validation}(c-d) where the (111) reflection was probed. We refer to this one as Sample 2.

The two datasets were collected under comparable optical configurations to assess both reconstruction performance and model transferability. Both samples were single crystals of industrially pure aluminum (99.99\%), with dimensions $0.7 \times 0.7 \times 10\,\mathrm{mm}^3$, oriented with the long $[1\bar{1}0]$ axis perpendicular to the scattering plane. Prior to the experiments, the samples were annealed for 10~h at 590~$^{\circ}$C in atmospheric air and slowly cooled in the furnace to minimize residual strain. All DFXM measurements were performed using 17 keV photons selected by a Si(111) Bragg–Bragg double-crystal monochromator. The incident beam was vertically focused using a compound refractive lens (CRL) consisting of 58 one-dimensional Be lenslets with a radius of curvature $R = 100\,\mu$m, yielding an effective focal length of 72 cm. The resulting beam profile at the sample was approximately line-focused, with a height of about 500 nm (FWHM) in the vertical direction and several hundred microns in the horizontal direction. This horizontal line beam illuminated a single plane within the crystal, thereby defining the microscope observation plane (Fig.\ref{dfxm}). Crystal alignment into the Bragg condition was performed using a near-field camera positioned 40 mm behind the sample. After alignment, the near-field detector was removed and the diffracted beam was magnified using an X-ray objective lens composed of 88 Be parabolic lenslets (two-dimensional focusing optics), each with $R = 50\,\mu$m. The entry plane of the imaging CRL was positioned approximately 275-280 mm downstream of the sample along the diffracted beam path. The magnified image was projected onto a far-field detector positioned approximately 5 m from the sample. The far-field detector used an indirect detection scheme consisting of a scintillator, visible-light microscope optics, and a PCO.edge sCMOS camera. The visible optics could switch between 2$\times$ and 10$\times$ objectives. The present analysis focuses on the highest magnification configuration (10$\times$ visible objective), corresponding to an effective pixel size of approximately 40 nm along the laboratory $y$ direction and approximately 120 nm along the laboratory $x$ direction. Lower magnification (2$\times$) configurations correspond to proportionally larger effective pixel sizes. 
\par In this work, we focused on rocking curve scans in both datasets. Rocking scans were acquired by rotating the tilt angle $\phi$ (see Fig.~1) about the laboratory $y$ axis over a defined angular range sampled in 30 steps. These one-dimensional scans of the rocking curve map components of the displacement gradient tensor field, including local strain and misorientation, enabling visualization of dislocations \cite{Yildirim2023}. For the 3D reconstruction shown in Fig.~\ref{3d} (Sample 1), rocking scans were collected for 301 $z$-layers with 1~$\mu$m spacing between successive slices. The rocking curve measurements were done over an angular range of $\degree$ with $\degree$ per step for Sample 1, totaling 31 images in the rocking curve. For sample 2, rocking curves were collected over a range of 0.02 $\degree$ with 0.002$\degree$ per step. For both experiments, the exposure time was 1 seconds per frame. The resulting four-dimensional dataset $(x, y, z, \phi)$ was imported for analysis. For further details see Refs.~\citeonline{Yildirim2023, Borgi2025}.



\subsection*{Lightweight Convolutional neural networks}

Convolutional Neural Networks (CNNs)~\cite{lecun} are widely used deep learning models designed for image tasks such as classification, object detection, and segmentation. Unlike traditional neural networks where every neuron is connected to all others, CNNs use convolution operations to focus on small regions of an image, allowing the network to learn patterns like edges, textures, and shapes more efficiently. This structure reduces the number of parameters and improves performance.

A typical CNN is made up of several key layers:
\begin{itemize}
    \item \textbf{Convolutional layers}, which apply filters to extract local features from the image.
    \item \textbf{Activation functions}, such as ReLU, which introduce non-linearity and help the model learn complex patterns.
    \item \textbf{Pooling layers}, which reduce the size of feature maps, speeding up computations and helping prevent overfitting.
    \item \textbf{Fully connected layers}, which take the learned features and output the final classification.
\end{itemize}

Standard convolutional layers use 3D filters that operate across both spatial dimensions and the full channel depth. While effective, this is computationally expensive because each filter processes all input channels simultaneously.

To improve efficiency, particularly in mobile or real-time applications, \textbf{lightweight CNNs} often use \textit{depthwise separable convolution}, which factorizes a standard convolution into two steps:
\begin{enumerate}
    \item \textbf{Depthwise convolution} ,  applies a spatial filter to each input channel independently, capturing spatial patterns without mixing channels.
    \item \textbf{Pointwise convolution} ,  applies a $1 \times 1$ convolution across channels to combine their information and learn inter-channel dependencies.
\end{enumerate}

This separation reduces computational cost by avoiding large filters that span all channels at once. Further, a standard convolution with kernel size $k \times k$, $C_{\mathrm{in}}$ input channels, and $C_{\mathrm{out}}$ output channels, the number of parameters is:
\begin{equation}
    k \times k \times C_{\mathrm{in}} \times C_{\mathrm{out}}.
\end{equation}

In a depthwise separable convolution, the parameter count becomes:
\begin{equation}
    k \times k \times C_{\mathrm{in}} + C_{\mathrm{in}} \times C_{\mathrm{out}},
\end{equation}
where the first term is from the depthwise convolution and the second from the pointwise convolution. This factorization greatly reduces parameters and computation.

CNNs are usually trained using the backpropagation algorithm\cite{Rojas1996}, which iteratively adjusts network weights to minimize classification errors. The training process involves two main steps: forward propagation and backward propagation. In forward propagation, input data passes through the network, generating predictions. The loss function then quantifies the discrepancy between predicted and true labels.

One commonly used loss function for classification tasks is the cross-entropy loss, defined as:
\begin{equation}
\mathcal{L} = -\frac{1}{N} \sum_{i=1}^{N} \sum_{j=1}^{C} y_{i,j} \log(\hat{y}_{i,j}),
\end{equation}
where $N$ is the number of training samples, $C$ is the number of classes, $y_{i,j}$ is a binary indicator that equals 1 if the true class for sample $i$ is $j$, and $\hat{y}_{i,j}$ is the predicted probability for class $j$.
The cross-entropy loss penalizes incorrect predictions, particularly when the model assigns low probability to the correct class. This makes the model produce probability distributions that align closely with true labels.

During backpropagation, the gradient of the loss function is computed with respect to each weight in the network using the chain rule of differentiation. These gradients are then utilized by optimization algorithms such as Stochastic Gradient Descent (SGD) \cite{Gardner1984} or Adaptive Moment Estimation (Adam)\cite{adam} to update the weights, minimizing the loss iteratively over multiple training epochs.

\section*{Data Availability}

The codes for this study are available and accessible via the
link https://github.com/Benhadjira/DL-WBDF


\section*{Author Contributions Statement}
A.B. developed the machine learning model, performed the formal analysis, and processed the data. The research idea was conceived by C.D., S.B., V.F.N., and C.Y. Experimental measurements were carried out by C.Y. The initial manuscript draft was written by A.B. with input from C.Y. All authors reviewed and approved the final version of the manuscript.

\section*{Competing Interests Statement}
The authors declare no competing interests.

\section*{Acknowledgement}
We thank the ESRF for providing beamtime at ID06-HXM and ID03. C.Y. acknowledges financial support from the ERC Starting Grant “D-REX” No. 101116911. S.B. acknowledges support from ERC Advanced Grant
nr. 885022 and from the Danish ESS lighthouse on hard materials in 3D, SOLID. We also thank Prof. L. Dresselhaus-Marais for providing the samples and for the fruitful discussions.

\section*{Funding information}

This project has been partly funded by the European Union’s
Horizon 2020 Research and Innovation Programme under the
Marie Sklodowska-Curie COFUND scheme with grant
agreement No. 101034267 and the European Research
Council (ERC) under the European Union’s Horizon 2020
Research and Innovation Programme (grant agreement No.
10116911)


\bibliography{ref}

\end{document}